\documentclass[12pt]{iopart}

\RequirePackage{lineno}
\usepackage{graphicx}
\usepackage{dcolumn}
\usepackage{epstopdf}
\usepackage{color}
\expandafter\let\csname equation*\endcsname\relax
 \expandafter\let\csname endequation*\endcsname\relax
\usepackage{amsmath}
\usepackage{amssymb}
\usepackage{bm}

\usepackage{subfigure}
\usepackage{hyperref}
\definecolor{dgreen}{cmyk}{1.,0.,1.,0.2}        
\definecolor{orange}{cmyk}{0.,0.353,1.,0.}    

\def\bea {\begin{eqnarray}}
\def\eea {\end{eqnarray}}
\def\ra {\rightarrow}
\def\nn {\nonumber}
\def\l {\left(}
\def\r {\right)}
\def \la{\langle}
\def \ra{\rangle}

\newcommand{\muB}{$\mu_{\rm B}$}

\newcommand{\sNN}{$\sqrt{s_{\rm NN}}$}

\newcommand{\CBS}{$C_{BS}$}
\newcommand{\CQS}{$C_{QS}$}
\newcommand{\CQB}{$C_{QB}$}

\newcommand{\acceptance}{$|\eta|<0.5$ and $0.2 < p_{\rm T}< 2.0$~GeV/c}

\begin{document}


\title{Diagonal and off-diagonal susceptibilities of conserved quantities in relativistic
  heavy-ion collisions}

\author{Arghya Chatterjee$^1$, Sandeep Chatterjee$^{2,1}$, 
Tapan K. Nayak$^1$ and Nihar Ranjan Sahoo$^3$}
\address{$^1$Variable Energy Cyclotron Centre, HBNI, Kolkata 700064, India}
\address{$^2$School of Physical Sciences, National Institute of
  Science Education and Research, Jatni, 752050, India}
\address{$^3$Texas A\&M University, College Station, Texas 77843, USA}

\ead{arghya@vecc.gov.in}

\vspace{10pt}
\date{ \today}

\begin{abstract}

Susceptibilities of conserved quantities, such as baryon number, strangeness 
and electric charge are sensitive to the onset of quantum
chromodynamics (QCD) phase transition, and are
expected to provide information on the matter 
produced in heavy-ion collision experiments. A comprehensive study of
the second order diagonal susceptibilities and cross correlations has
been made within a thermal model approach of the hadron resonance gas
(HRG) model as well as with a hadronic transport model,
UrQMD. We perform a detailed analysis of the effect of detector acceptances 
and choice of particle species in the experimental measurements of the
susceptibilities for heavy-ion collisions corresponding to 
\sNN~=~4~GeV to 200 GeV. The transverse momentum cutoff dependence of 
suitably normalised susceptibilities are proposed as useful observables 
to probe the properties of the medium at freezeout.

\end{abstract}



\section{Introduction}

Heavy-ion collisions at relativistic energies create matter at
extreme conditions of temperature and energy density, consisting of
deconfined quarks and gluons. The major motivations of
facilities at the Relativistic Heavy Ion Collider (RHIC) at Brookhaven National
Laboratory, the Large Hadron Collider (LHC) at CERN and FAIR at GSI
are to study the formation of this new form of matter, called quark-gluon plasma (QGP) and study its
basic properties. 
In the Quantum Chromodynamics (QCD) phase
diagram, location of the QCD critical point could be one of the compelling
discovery in the heavy-ion collisions. The end point of the first order
phase transition between hadronic matter to QGP phase is known as QCD
critical point, after which there is no genuine phase transition but a
cross over from hadronic to quark-gluon degrees of freedom~\cite{QCDPhasediag1,QCDPhasediag2}.
While the temperature for this crossover transition at 
zero baryon chemical potential is now theoretically established from lattice QCD 
(LQCD) computations~\cite{LQCD}, the existence and location of the
critical point at 
non-zero baryon density is far from settled~\cite{Raja,Gavai}. On the experimental 
front, the ongoing beam energy scan program at RHIC is attempting to
establish the onset of phase transition and locate the critical point.

One of the foremost methods for the critical point search is through 
measurements of fluctuations of conserved quantities~\cite{Sus1,Sus2}. 
These quantities have been estimated theoretically - both on the lattice~\cite{LQCD1, LQCD2,
LQCD3, LQCD4, LQCD5} as well as in models~\cite{model1,model2,model3,model4,model5,model6,
model7,model8,model9,model10,Sumit}. Recently, RHIC experiments 
have published results on the diagonal susceptibilities of conserved charges, like, net electric 
charge~\cite{STAR_netQ,PHENIX_netQ} and net
baryon~\cite{STAR_netP}. Comparison of the experimental data to HRG model
calculation~\cite{HRGsus} shows that at higher beam energies
(\sNN~$>62.4$~GeV) there is reasonably good agreement between data and 
the model. This suggests that at these energies, 
the observed fluctuation of conserved charges must be grand-canonical thermal fluctuations. 
However, at lower energies there is disagreement between data and model
calculation for certain conserved charge 
susceptibilities indicating non-thermal fluctuations. It is not yet clear whether such non-thermal 
behaviour is a signature of QCD critical physics.

Similar to the diagonal susceptibilities, 
different combinations of higher order off-diagonal susceptibilities of net charge, net baryon 
number and net strangeness can be estimated using different combinations of higher order central 
moments of conserved charges in the experiments. The measurement of these observables give 
information to explore the flavor carrying susceptibilities and also the nature of QCD phase 
transitions~\cite{Sus3,Majumder,Gavai2,deepak}.

There are several constraints in the experimental measurements which need to be 
properly understood in order to interpret the results and compare with theoretical 
calculations. 
A limited set of produced particles are measured 
on an event-by-event basis. The neutron, whose contribution to 
net baryon ($N_{\rm B}$) is as good as the proton ($p$), is not measured. 
$\Lambda$-baryon is the lightest strange baryon and hence contributes most 
significantly to the baryon-strangeness correlation. Although $\Lambda$-baryon is measured 
over an ensemble through its charged daughter particles, its
measurement on an event-by-event level poses a daunting task.
Thus, only the moments of net charge ($N_Q$) 
are measured faithfully. For the other charges we have to rely on proxies, 
e.g. net proton serves as a good proxy for net baryon and net kaons
serve as a proxy for net strangeness.

We have studied the dependence on the observed particle sets for all the second order 
susceptibilities. 
We have made a comprehensive study of the effect due to finite number of the conserved quantities. 
Ideally, in order to observe grand canonical fluctuations of the conserved 
charges, the fraction $\mathcal{R}$ of the conserved charge carried by the 
system to that of the total available charge carried by the bath as well as the system should 
be much smaller than half ($\mathcal{R}<<0.5$). However, in 
reality for full overlap collisions, the $N_{\rm B}$ equals to 
twice the mass number of the nucleus and $N_{\rm Q}$ equals to twice the atomic number 
of the nucleus that are distributed in the momentum rapidity ($y$) direction 
within $|y_{\text{max}}|\sim\log\left(\text{\sNN}/m_{\rm p}\right)$. Thus, the final 
distribution of this $N_B$ and $N_Q$ into the system and bath depends on three 
factors: (a) the available experimental acceptance, (b) \sNN, and (c) the baryon stopping 
phenomenon which decides the initial rapidity distribution of the conserved 
charges. It is important to note that the strong \sNN~dependence of baryon 
stopping rubs off on to $\mathcal{R}$, complicating the comparison of measured 
moments at different \sNN.

In this article, we have used the Ultra-relativistic Quantum Molecular
Dynamics (UrQMD v3.3) as well as the hadron resonance gas (HRG) model to
analyze the diagonal and off-diagonal susceptibilities of the conserved
charges. Model descriptions for UrQMD and HRG are presented in Section II. 
In Section III, we discuss the observables used for the present study.
In Section IV, we present the collision energy dependence of all the
susceptibilities. We discuss the species dependence on the
observables and effect of detector acceptances. We present the results
from both the models. The article is summarised in Section V. An
appendix at the end discusses the estimation of statistical errors
associated with the observables.

\section{Model considerations}

UrQMD is a microscopic transport model~\cite{UrQMD1,UrQMD2}. In this
model, the 
space-time evolution of the fireball is studied in terms of excitation 
of color strings which fragment further into hadrons, the covariant 
propagation of hadrons and resonances which undergo scatterings and finally 
the decay of all the resonances. This model setup has been quite successful 
and widely applied in heavy-ion phenomenology~\cite{UrQMD1,UrQMD2}. It 
has also been previously used to compute several 
susceptibilities~\cite{UrQMDapply1,UrQMDapply2,Bleichersus,Sahoo,Bhanu}. The acceptance window plays 
an important role in such studies. The initial distribution of $N_B$ and 
$N_Q$ in $y$ is a consequence of the baryon stopping phenomenon which has 
a strong \sNN~ dependence- as a result at higher \sNN~the mid-rapidity 
region is almost free of $N_B$ and $N_Q$ while at lower \sNN~almost all the 
$N_B$ and $N_Q$ are deposited in the mid-rapidity region. This is also 
expected to have significant effect on the fluctuations of conserved quantities. 
This \sNN~ dependent baryon stopping phenomenon is dynamically included 
in the UrQMD approach. Here, we have generated around a million events 
per beam energy from \sNN~$=4-200$ GeV. 

We have compared the UrQMD results with those from a thermal 
approach, by using the HRG model. Here it is important to note that while the 
HRG results are expected to reflect the susceptibilities at the time of chemical 
freezeout where inelastic collisions cease, the UrQMD results reveal the above 
quantities at the kinetic freezeout surface where the elastic collisions cease. Thus, 
while the latter are more realisitic in the context of heavy ion collisions, the 
HRG results here serve as useful guide for a qualitative understanding 
of the results. The HRG model consists of all hadrons and resonances as listed in the 
Particle Data Book~\cite{PDG} within the framework of a multiple species non-interacting 
ideal gas in complete thermal and chemical equilibrium. The only parameters 
are the temperature $T$, chemical potentials $\mu_B$, $\mu_Q$ and $\mu_S$ 
corresponding to the conserved quantities of baryon number $B$, electric 
charge $Q$ and strangeness $S$, and the volume $V$ of the fireball which 
are obtained by fits to data. The HRG model has been found to provide a 
very good description of the mean hadron yields using a few thermodynamic 
parameters at freeze-out (for a recent compilation of the freeze-out parameters, 
see Ref.~\cite{HRGyield}). In this work, we have used the same 
parametrisation for the \sNN~dependence of $T$, $\mu_B$, $\mu_Q$ and $\mu_S$ 
as given in Ref.~\cite{model8}. The chemical freezeout volumes, $V\l\text{\sNN}\r$ are taken from 
Ref.~\cite{HRGyield}. These parameters are listed in Table~\ref{tab.volume}.

\begin{table}[t]    
\begin{center}
\begin{tabular}{ |c|c| c| c| c| c |} 
\hline
\sNN~(GeV) & $10^4 V$ ($\text{MeV}^{-3}$) & $T$ (MeV) & $\mu_B$ (MeV)
  & $\mu_Q$ (MeV)& $\mu_S$ (MeV) \\
\hline
6.27	&	1.4 & 130.8 & 482.4 & 12.7 & 106.5 \\
7.62	&	1.3 & 139.2 & 424.6 & 11.7 & 96.1 \\
7.7	&	1.3 & 139.6 & 421.6 & 11.6 & 95.5 \\
8.76	&	1.1 & 144.2 & 385.7 & 10.9 & 88.8 \\
11.5	&	1.5 & 151.6 & 316.0 & 9.5  &  75.0\\
17.3	&	2.0 & 158.6 & 228.6 & 7.4  &  56.5\\
39	&	1.7 & 164.2 & 112.3 & 4.1 &  29.4\\
62.4	&	1.8 & 165.3 &  72.3 &  2.8 &  19.4\\
130	&	2.1 & 165.8 &  35.8 &  1.4 &  9.8\\
200	&	2.5 & 165.9 &  23.5 &  1.0 &  6.4 \\   
\hline 
\end{tabular}
\end{center}
\caption{The chemical freezeout parameters extracted from mid-rapidity data at different 
\sNN~\cite{model8,HRGyield}.}
\label{tab.volume}
\end{table}

Lately, susceptibilities have also been employed 
to study the freeze-out conditions of the fireball within the HRG model~\cite{HRGsus}. 
All the quantities of interest can be computed from the partition 
function $Z\l V,T,\mu_B,\mu_Q,\mu_S\r$
\bea
\ln &Z&= \sum_i\ln Z_i\nonumber \\
&& = \sum_i \frac{aVg_i}{\l2\pi\r^3}d^3p\ln\l 1+ae^{\l-\l p^2+m_i^2\r+\mu_i\r/T}\r\\
&& =VT^3\sum_i\frac{g_i}{2\pi^2}\left(\frac{m_i}{T}\right)^2 
 \sum_{l=1}^{\infty}\left(-a\right)^{l+1}l^{-2}K_2\left(lm_i/T\right) 
   \nonumber \\
&& \exp[l\left(B_i\mu_B+Q_i\mu_Q+S_i\mu_S\right)/T],
\label{eq.ZHRG}
\eea
where $a=-1$ for mesons and $1$ for baryons, $g_i$, $m_i$, $B_i$, $Q_i$, 
$S_i$ and
\bea
\mu_i = B_i\mu_B+Q_i\mu_Q+S_i\mu_S
\label{eq.hadronmu}
\eea
refer to the degeneracy factor, mass, baryon number, electric charge, 
strangeness and hadron chemical potential respectively of the $i$th 
hadron species, $V$ is the part of the fireball volume under study 
that can also be called the system and $K_2$ is the modified Bessel 
function of the second kind. From $\ln Z$, all thermodynamic quantities could be computed.

\section{Observables and methods}

The susceptibilities of the conserved quantities of the strongly interacting 
matter in thermal and chemical equilibrium can be computed within the 
grand canonical ensemble (GCE) from partial derivatives of the 
pressure~($P$) with respect to the chemical potentials 
\bea 
 \chi^{ijk}_{BQS} &=& 
 \frac{\partial^{i+j+k} \l P/T^4\r}{\partial^i\l\mu_B/T\r\partial^j 
 \l\mu_Q/T\r\partial^k\l\mu_S/T\r}\nn,\\
 \label{eq.sus}
\eea 
where the $P$ is obtained from $\ln Z$ as follows 
\bea 
P=\frac{T}{V}\ln Z 
\label{eq.PHRG}
\eea 

From the experimental point of view it is straightforward to compute the 
central moments, $\rm{M}$,
\bea 
 \rm{M}^{ijk}_{BQS} &=& \la\l B-\la B\ra\r^i\l Q-\la Q\ra\r^j\l S-
\la S\ra\r^k\ra. 
 \label{eq.mom}
\eea 
Using the fact that the generating function for the cumulants is given 
by the logarithm of that of the moments, one can express one in terms 
of the other. Up to the second order, this relationship is one-to-one,
such as,
\bea 
\chi^{11}_{XY} =\frac{1}{VT^{3}} \rm{M}^{11}_{XY}. 
\label{eq.sustomom}
\eea 
Using this relation, all the diagonal and non-diagonal 
susceptibilities of second order can be expressed in terms of
second order central moments ($\sigma$):
\begin{center}$\begin{pmatrix}
   \sigma_{Q}^{2}       & \sigma_{QB}^{1,1} & \sigma_{QS}^{1,1} \\ \\
   \sigma_{BQ}^{1,1}       & \sigma_{B}^{2} & \sigma_{BS}^{1,1} \\ \\
   \sigma_{SQ}^{1,1}       & \sigma_{SB}^{1,1} & \sigma_{S}^{2}
\end{pmatrix}$\end{center}

The ratios of of $\chi^{ij}_{XY}$ to 
$\chi^2_Y$ can be suitably constructed to cancel the volume effect.
In the quasiparticle picture of quarks and 
gluons, the ratios $\chi^{11}_{BS}/\chi^{2}_{S}$ and $\chi^{11}_{QS}/\chi^{2}_{S}$ 
become -1/3 and 1/3 respectively. It is not possible to find such simple 
factors for other ratios like $\chi^{11}_{QB}/\chi^2_B$ that receive contribution 
from both light and strange quarks. 
Thus the following the ratios of off-diagonal and 
diagonal susceptibilities have been constructed:
\bea 
 C_{BS} = -3\frac{\chi^{11}_{BS}}{\chi^2_S},&~~~~&  C_{SB} = -\frac{1}{3}\frac{\chi^{11}_{BS}}{\chi^2_B}, \\
 C_{QS} = 3\frac{\chi^{11}_{QS}}{\chi^2_S},&~~~~&  C_{SQ} = \frac{\chi^{11}_{QS}}{\chi^2_Q}, \\
 C_{QB} = \frac{\chi^{11}_{QB}}{\chi^2_B},&~~~~&  C_{BQ} = \frac{\chi^{11}_{QB}}{\chi^2_Q}.
\eea 
We have studied the susceptibilities and these ratios as a function of collision energy and
detector acceptances in terms of pseudorapidity~($\eta$) and
transverse momentum~($p_T$).

\section{Beam energy dependence}\label{sec.beam}

 \begin{figure}[tbp]
\centering 
  \includegraphics[width=0.43\textwidth]{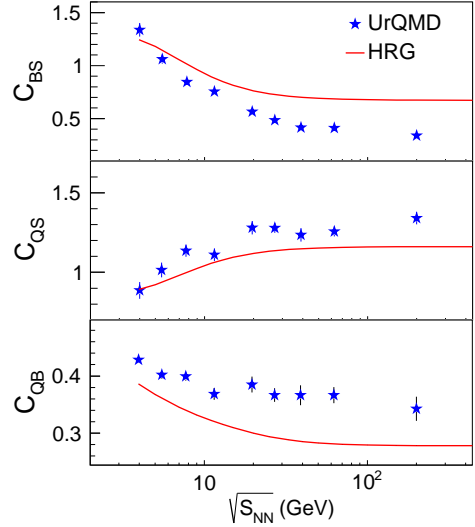}
\caption{(Color online) Beam energy dependence of ratios of off-diagonal to 
diagonal susceptibilities for central (0-5\%) Au+Au collisions using
UrQMD and HRG models.}
\label{fig.all} 
\end{figure}

In Figure~\ref{fig.all}, we present the ratios, $C_{BS}$, $C_{QS}$, and
$C_{QB}$ as a function of  \sNN~for top central (0-5\% of total cross section) Au+Au 
collisions using both UrQMD and HRG models.
In this construction, the ratios are calculated by considering 
all the charges, baryons and strange
particles within \acceptance~. 
The trends of the ratios
for both the models are similar, although there are quantitative
differences.
In going from low to high collision energy, the values of
\CBS~decrease and remain constant after \sNN~=~27GeV, whereas the
values of \CQS~increase and then remain constant after
\sNN~=~20GeV. The values of \CQB~show decreasing trend with increasing \sNN. 

These trends can be understood on the basis of the thermal 
model framework of the HRG. $C_{BS}$ gets dominant contribution from $\Lambda$ 
(being the lightest strange baryon) in the numerator while in the denominator, 
it gets contribution mainly from the kaons (being the lightest strange mesons). 
With decreasing \sNN, \muB~increases which in turn enhances the relative 
contribution from $\Lambda$ compared to kaons resulting in the increasing trend 
of $C_{BS}$.  In case of $C_{QS}$, it receives dominant contribution
from the kaons both in the numerator and 
denominator. However the contribution from the lightest strange baryon $\Lambda$ 
to $\chi^2_S$ keeps growing with decreasing \sNN. This 
results in the monotonic increasing trend for $C_{QS}$ with \sNN. Finally, 
the weak rising trend of $C_{QB}$ with decreasing \sNN~can be traced to the 
contribution from the multiply charged $\Delta$ to the numerator.

\subsection{Species Dependence of the hadrons}\label{subsec.hadronchoice}

\begin{figure*}[tbp]
\centering 
  \includegraphics[width=0.9\textwidth]{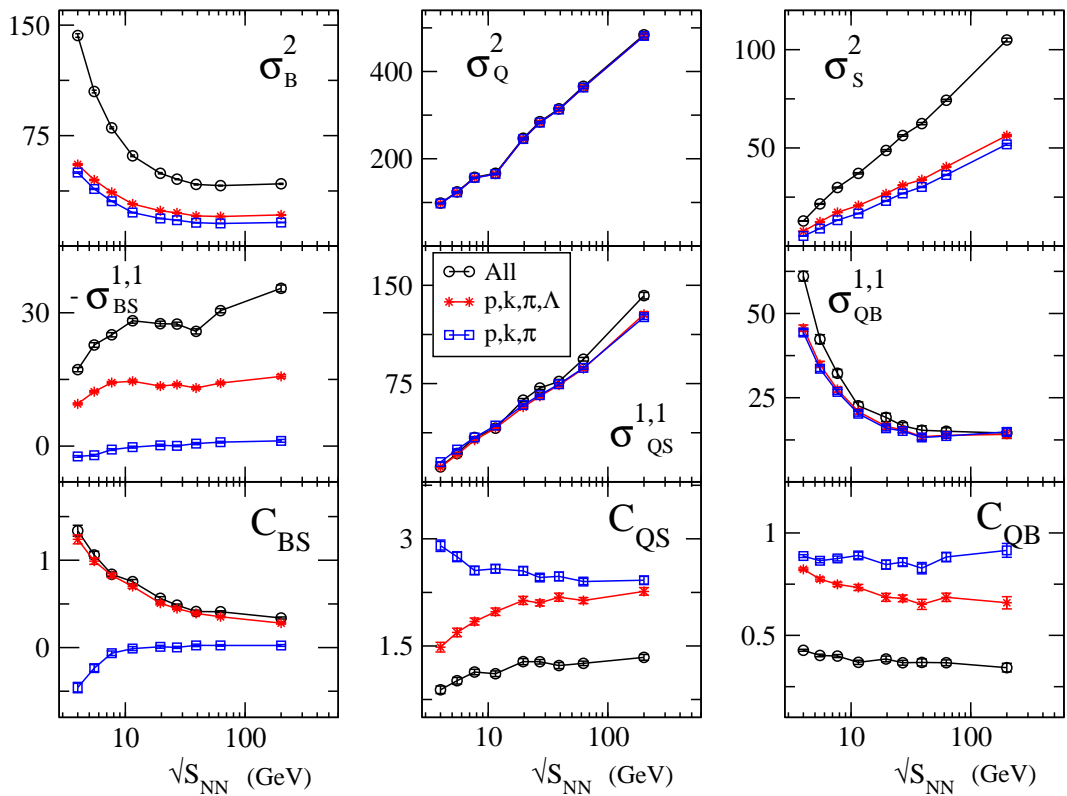}
\caption{(Color online) Particle species dependence on the diagonal and 
off-diagonal susceptibilities and their ratios in central (0-5\%) Au+Au 
collisions using the UrQMD model calculations.}
\label{fig.hadronchoice} 
\end{figure*}

Susceptibilities and their ratios have a strong species dependence. 
In the experiments, only charged hadrons are measured. At present, it is possible 
to perform an event-by-event analysis of only $\pi^+$, 
$K^+$, $p$ and their anti-particles. Neutral hadrons, like $K^0$, $n$ and 
$\Lambda$, which contribute significantly to the strangeness fluctuation, baryon 
fluctuation and baryon-strangeness correlation, respectively, are
not measured on an event-by-event basis.
We have estimated the effect of such missing contributions by computing all the 
ratios for three different hadron sets: (i) considering all hadrons as in
Fig.~\ref{fig.all}, (ii) with  $\pi^\pm$, $K^\pm$, $p$, $\bar{p}$ and
$\Lambda$ and $\bar{\Lambda}$, and (iii) with  only $\pi^\pm$, $K^\pm$, $p$
and $\bar{p}$.
The results of the study using UrQMD model are 
shown in Fig.~\ref{fig.hadronchoice} as a function of collision
energy. Ratios like $C_{BS}$, $C_{QS}$ and $C_{QB}$ are constructed using variances 
(like $\sigma^2_B$, $\sigma^2_S$ and $\sigma^2_Q$), and covariances (like $\sigma^{1,1}_{BQ}$, 
$\sigma^{1,1}_{SB}$ and $\sigma^{1,1}_{QB}$) of different conserved quantities. Large 
differences have been observed for the three cases. In all cases, it is clear
that using only $\pi^\pm$, $K^\pm$, $p$, $\bar{p}$ as event-by-event measurements, 
the sensitiveness of the ratios as a function of collision energy reduces to a large 
extent. For $C_{BS}$, there is practically no difference between the cases with all 
particles and those with $\pi^\pm$, $K^\pm$, $p$, $\bar{p}$ and $\Lambda$ and 
$\bar{\Lambda}$. But the trend for only $\pi^\pm$, $K^\pm$, $p$, $\bar{p}$ is very 
different. It is necessary to analyse at least $\Lambda$ and $\bar{\Lambda}$ included 
in the measurements. Here we note that an additional hurdle in experiments 
will be to distinguish $\Sigma_0$ from $\Lambda$. However, our results here with all 
the particles as well as only $\pi^\pm$, $K^\pm$, $p$, $\bar{p}$ and $\Lambda$ and 
$\bar{\Lambda}$ are almost identical. Thus the role of $\Sigma_0$ is probably sub-dominant. 
It is well known that in UrQMD, the yields of multi-strange baryons are highly 
underestimated ~\cite{SteinPRC, SteinIOP}. It might be possible that similar reason also leads to the above 
observation of marginal influence of the multi-strange baryons towards strange 
susceptibilities as well as baryon-strange correlations.
$C_{QB}$ on the other hand is almost constant across all \sNN- the case with only 
$\pi^\pm$, $K^\pm$, $p$ and $\bar{p}$ is roughly twice of the case when we include 
all hadrons due to the missing neutrons in the denominator in the former.

 \begin{figure}[tbp]
\centering 
  \includegraphics[width=0.43\textwidth]{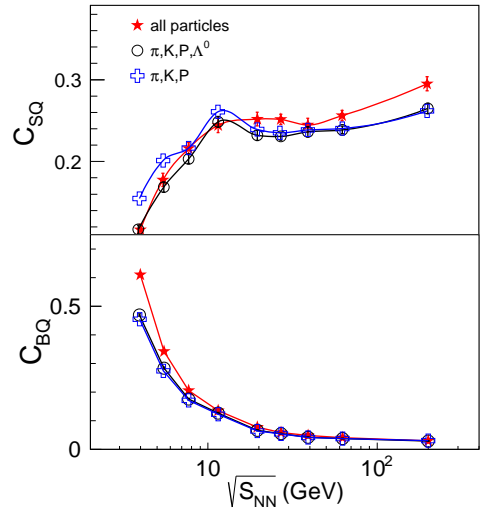}
\caption{(Color online) The measurements for $C_{SQ}$ and $C_{BQ}$ are robust under 
the different choices of particle set for central (0-5\%) Au+Au collisions}. 
\label{fig.hadronchoicerev} 
\end{figure}

In order to reduce the dependence on the choice of hadron set we also look at 
the other ratios, $C_{SQ}$ and $C_{BQ}$, where only charged hadrons contribute.
We show the results of these two quantities from UrQMD in Fig.~\ref{fig.hadronchoicerev}. 
For both of these quantities, the leading contributors to numerator as 
well as denominator come from the measured hadron sets $\pi^\pm$, $K^\pm$, $p$ and $\bar{p}$. 
Thus, we find that these results are quite stable to further inclusion 
of other hadrons and hence can be estimated in experiments with the limited 
particle identification capability. 

\subsection{Acceptance dependence}\label{subsec.acc}

Limited event-by-event particle identification, realistic efficiency corrections in 
the experiments and finite kinematic acceptances in $p_T$ and $\eta$-
all of these contribute to 
dilute  the signal for susceptibilities in the experiments~\cite{KochBzdak}. Some of 
these effects have been already discussed earlier~\cite{issues1, issues2, model10, issues4, 
issues5, issues6}. In the previous section we discussed the effect of limited particle 
identification on susceptibilities. We will now discuss specifically the effect of the window of detector 
acceptances in terms of $\eta$ and $p_T$ windows for measuring the susceptibilities.

Ideally, grand canonical fluctuations trivially scale with system volume when in 
contact with an infinite bath. On the other hand, in heavy-ion collisions for large enough 
acceptance it is possible that the system size becomes comparable with that of the 
bath and even larger. Global charge conservation in such cases cause 
suppression of thermal fluctuation resulting in non-thermal fluctuations~\cite{NetB1,NetB2}. 
Thus, the interpretation of conserved charge fluctuations in terms of thermal and 
critical fluctuations is not straightforward in such cases. Another factor that adds 
to the above complication is the fact that baryon stopping is not constant across 
\sNN, resulting in completely different distributions of conserved charges in $\eta$ 
for different \sNN. 

\begin{figure}[tbp]
\centering 
  \includegraphics[width=0.45\textwidth]{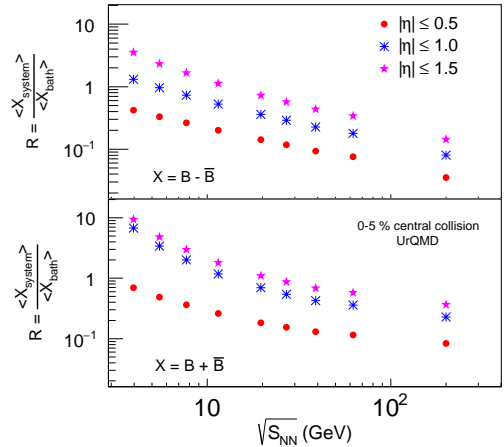}
\caption{(Color online) 
The ratios of net baryon (top panel) and total baryon (bottom panel) numbers within the system 
to the bath + system for central Au+Au collisions are plotted as a function of collision
energy using the UrQMD model. The results are presented for three
different $\eta$ windows.
} 
\label{fig:bathBysystem}
\end{figure}

The rapidity dependence of baryons affects the beam energy dependence 
of the ratio $\mathcal{R}$ of the number of baryons carried by the
system to the total carried by the system as well as bath. 
Ideally, in order to observe grand canonical
fluctuations $\mathcal{R}<<0.5$. We study this ratio for net baryons and
total number of baryons as a function of collision energy for central
Au+Au collisions using UrQMD model. The ratios are plotted for three
different $\eta$ windows (0.5, 1.0 and 1.5)
in Fig.~\ref{fig:bathBysystem}. 
We find that at high \sNN, $\mathcal{R}$ is sufficiently less than $0.5$. However, 
at lower \sNN, the $|\eta|<0.5$ case stays around $0.25$ while for even larger 
acceptance windows, $\mathcal{R}>0.5$ signalling the 
inapplicability of the grand canonical ensemble for $N_B$ fluctuations. Thus, a 
fixed $\eta$ window across all beam energies does not correspond to same system to 
bath effective volume ratio for all \sNN~\cite{sourendu}. This suggests that for 
low \sNN, $|\eta|<0.5$ is the upper bound for the acceptance window in
$\eta$. We have studied in detail the dependence of susceptibilities 
on the acceptance window in $p_T$ and $\eta$.

\subsubsection{Susceptibilities from HRG model\\}

\begin{figure*}[tbp]
\centering 
  \includegraphics[width=0.7\textwidth]{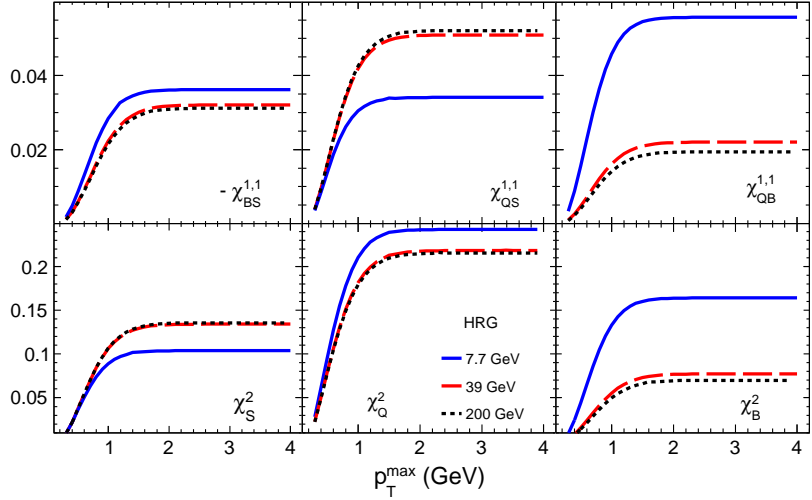}
\caption{(Color online) The ${p_T}_{\text{max}}$ dependence of the second order 
susceptibilities in the HRG model for central Au+Au collisions at
three colliding energies.}
\label{fig.pthrg} 
\end{figure*}

\begin{figure*}[tbp]
\centering 
  \includegraphics[width=0.7\textwidth]{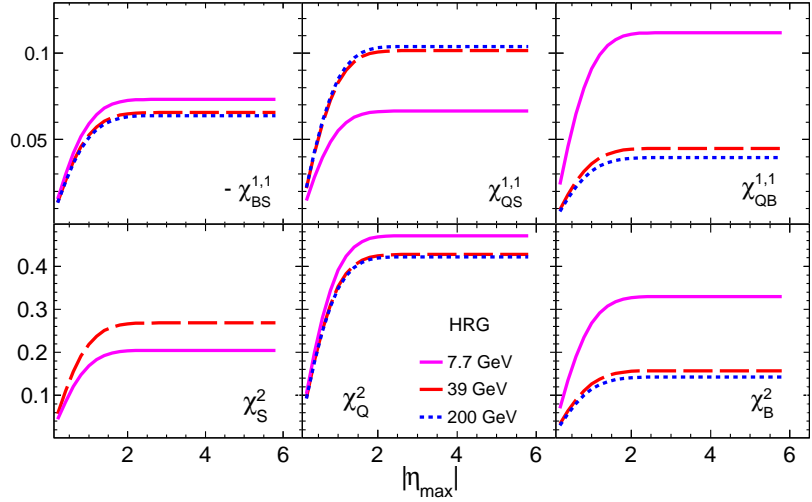}
\caption{(Color online) The $\eta_{\text{max}}$ dependence of the second order 
susceptibilities in the HRG model for central Au+Au collisions at 
three colliding energies. }
\label{fig.etahrg} 
\end{figure*}

Let us first analyse the acceptance dependence of the susceptibilities
of conserved quantities using the HRG model. 
From Eq.~\ref{eq.ZHRG} and \ref{eq.sus}, we can write the susceptibilities 
due to the $h$-th hadron as,
\bea
{\chi_h}^{ijk}_{BQS}&=&\frac{g_h}{\l2\pi\r^2}\sum_{l=1}^{\infty}
e^{l\mu_i/T}\l-a\r^{l+1}l^{\l i+j+k\r-4}B^iQ^jS^k\nonumber\\
&& \int_{-{y_r}_{max}}^{{y_r}_{max}}dy_r\text{Cosh}\l y_r\r
\int_{y_{min}}^{y_{max}}dy~ y^2 e^{-y \text{Cosh}\l y_r\r}\nonumber \\
&&
\eea
where $y_{min}=\frac{l}{T}\sqrt{{p_T}_{min}^2+m_h^2}$ and similarly for $y_{max}$. For 
ease of understanding, we have written down the 
explicit cutoff dependence on the momentum rapidity $y_r$ instead of the pseudorapidity 
$\eta$ which is more relevant experimentally. As noted from the integrand, the Boltzmann 
factor in terms of $y_r$ and transverse momentum $p_T$ is 
$e^{-l\sqrt{p_T^2+m^2}/T\text{Cosh}\l y_r\r}$. The $\text{Cosh}\l y_r\r$ dependence 
ensures that thermal production of particles is more strongly suppressed in $y_r$ as 
compared to $p_T/T$.

The susceptibilities have a strong dependence on the  
maximum value of transverse momentum (${p_T}_{\text{max}}$) 
value and
maximum $\eta$-window ($\eta_{\text{max}}$). 
These dependences are
studied for central Au+Au collisions for 
three collision energies and plotted in Figs.~\ref{fig.pthrg} and
\ref{fig.etahrg} as a function of (${p_T}_{\text{max}}$)  and
$\eta_{\text{max}}$, respectively. All the susceptibilities show
similar trends with both ${p_T}_{\text{max}}$ and $\eta_{\text{max}}$.
Small values of susceptibilities are observed for small ${p_T}_{\text{max}}$ and $\eta_{\text{max}}$
which steadily grow with increasing ${p_T}_{\text{max}}$ and $\eta_{\text{max}}$ as more 
phase space is included, finally saturating off to a constant value as the Boltzmann factor 
suppresses any further contribution from high $p_T$ and $\eta$.

\subsubsection{Susceptibilities from UrQMD model\\} 

\begin{figure*}[tbp]
\centering 
  \includegraphics[width=0.7\textwidth]{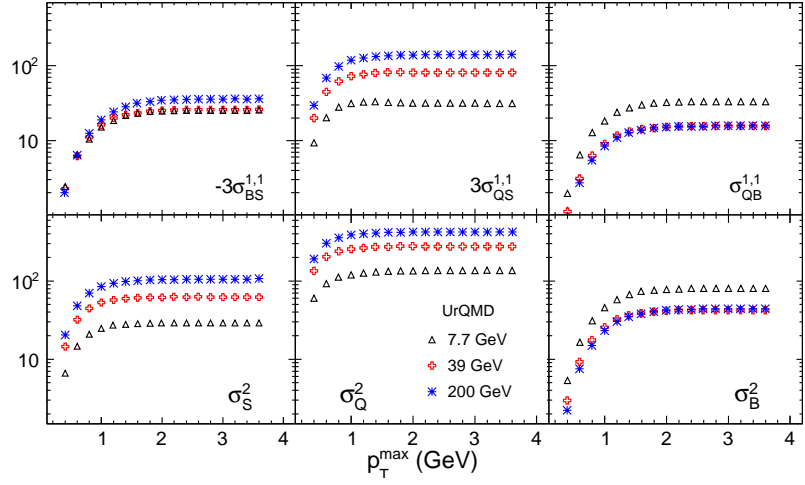}
\caption{(Color online) The ${p_T}_{\text{max}}$ dependence of the second order 
susceptibilities in the UrQMD model for central Au+Au collisions at
three colliding energies. }
\label{fig.pturqmd} 
\end{figure*}

\begin{figure*}[tbp]
\centering 
  \includegraphics[width=0.7\textwidth]{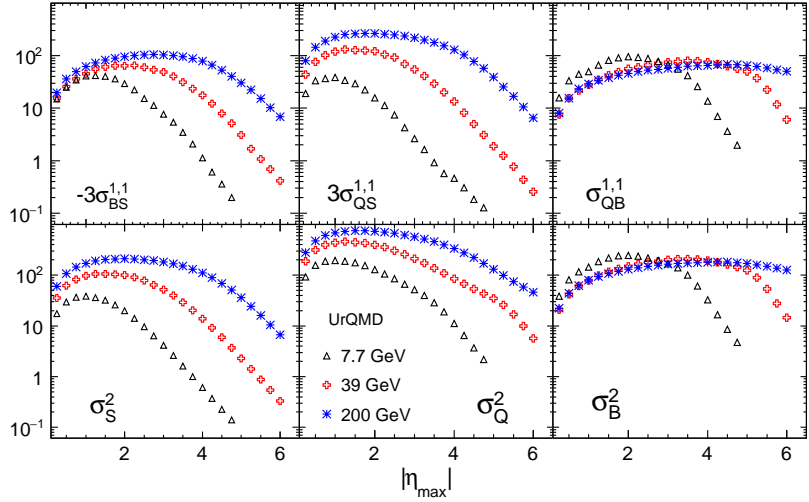}
\caption{(Color online) The ${\eta}_{\text{max}}$ dependence of the second order 
susceptibilities in the UrQMD model for central Au+Au collisions at
three colliding energies.}
\label{fig.etaurqmd} 
\end{figure*}

Figures~\ref{fig.pturqmd} and \ref{fig.etaurqmd} show the different second order 
susceptibilities with acceptances in terms of 
${p_T}_{\text{max}}$ and $\eta_{\max}$ as obtained using the UrQMD model.
The ${p_T}_{\text{max}}$ dependence of susceptibilities turns out to
be similar to those obtained from the HRG model. However, 
the $\eta_{\text{max}}$ dependence exhibits quite different behaviour,
primarily because of global conservation. There is 
an initial increase for small $\eta_{\text{max}}$ and attain maximum value 
at intermediate rapidity window within $\eta_{\max} \sim$ 1 to 2 units. At large 
rapidity window both variance and covariance terms go to zero because of the charge 
conservation effect at full phase space. This suggests that
$\Delta\eta \sim$ 2 to 3 capture the full essence of conserved charge correlations.

\subsubsection{Normalised susceptibilities from HRG and UrQMD\\} 

\begin{figure}[tbp]
\centering 
  \includegraphics[width=0.5\textwidth]{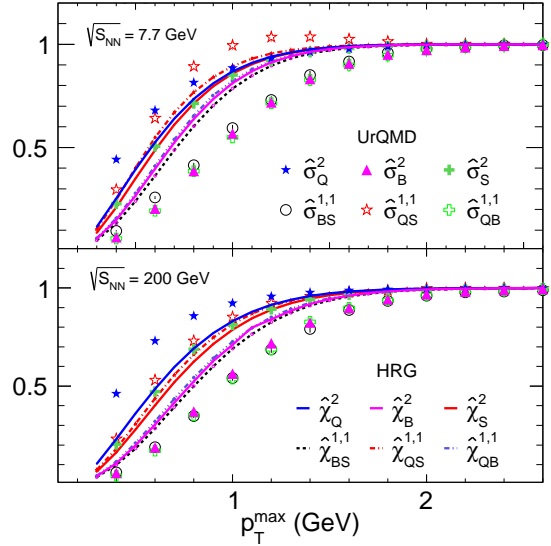}
\caption{(Color online) The ${p_T}_{\text{max}}$ dependence of all second order 
susceptibilities normalised by their values for ${p_T}_{\text{max}}=2$ GeV and 
$\eta_{\text{max}}=0.5$ for central Au+Au collisions at \sNN~=~7.7 GeV
(upper panel) and 200 GeV (lower panel) by using the HRG and UrQMD models.}
\label{fig.pt_all_onecanvas} 
\end{figure}

\vspace*{0.5cm}
The ${p_T}_{\text{max}}$ and $\eta_{\text{max}}$ dependence of the susceptibilities 
within the HRG and UrQMD models presented in the earlier sections can be compared and 
nicely summarised after they are suitably normalised. Here, we have normalised by 
their values at ${p_T}_{\text{max}}=2$ GeV and $\eta_{\text{max}}=0.5$. We denote these
normalised susceptibilities as $\hat{\chi}$ and $\hat{\sigma}$. Thus by 
construction, for ${p_T}_{\text{max}}=2$ GeV and $\eta_{\text{max}}=0.5$, $\hat{\chi}$ 
and $\hat{\sigma}$ are unity. Figure~\ref{fig.pt_all_onecanvas} 
shows ${p_T}_{max}$ dependence of the variance and covariance of conserved quantities, from both 
UrQMD and HRG models, in central Au+Au collisions for two collision energies.
For small ${p_T}_{\text{max}}$, the 
susceptibilities in both the models approach zero as the system phase space volume 
approaches zero. In all the cases the 
fluctuations grow with ${p_T}_{max}$ before saturating to a constant value. It
is interesting to observe a clear conserved charge ordering in these normalised susceptibilities
with the increase of ${p_T}_{max}$. This is observed in both HRG and
UrQMD model calculations. $\hat\chi^2_Q$ that receives contribution mainly from net pion reaches 
its saturation value fastest while $\hat\chi^2_B$ that gets contribution from net proton 
saturates at larger values of ${p_T}_{max}$. $\hat\chi^2_S$ which mainly gets contribution from kaons 
saturates at intermediate ${p_T}_{max}$, closer to that of $\hat\chi^2_Q$.

\begin{figure}[tbp]
\centering 
  \includegraphics[width=0.5\textwidth]{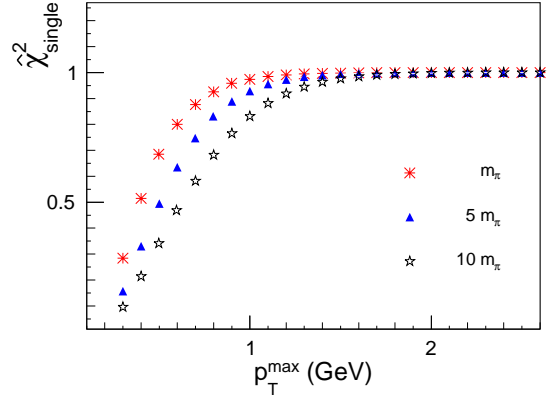}
\caption{(Color online) The ${p_T}_{\text{max}}$ 
dependence of the second order susceptibility 
normalised by the value for ${p_T}_{\text{max}}=2$ GeV and $\eta_{\text{max}}=0.5$ for a single 
particle system. The three curves are for three different masses of the particle.}
\label{fig.massordering} 
\end{figure}

In the HRG setup, it is easy to understand that such conserved charge ordering arises from 
the ordering of the masses of the hadrons that contribute dominantly to the different 
susceptibilities. This can be clearly understood within the framework of a 
single particle ideal gas. We have plotted the second order normalised susceptibility of this 
single particle ideal gas for three different masses of the particle. The results of 
these calculations are shown in Fig.~\ref{fig.massordering}. 
As the mass of the particle increases, the saturation in $\hat\chi^2_{single}$ kicks in 
at a higher ${p_T}_{max}$. As we go from HRG to UrQMD, we find the effect of the mass ordering 
gets even more pronounced as seen in Fig.~\ref{fig.pt_all_onecanvas}.

\begin{figure}[tbp]
\centering 
  \includegraphics[width=0.5\textwidth]{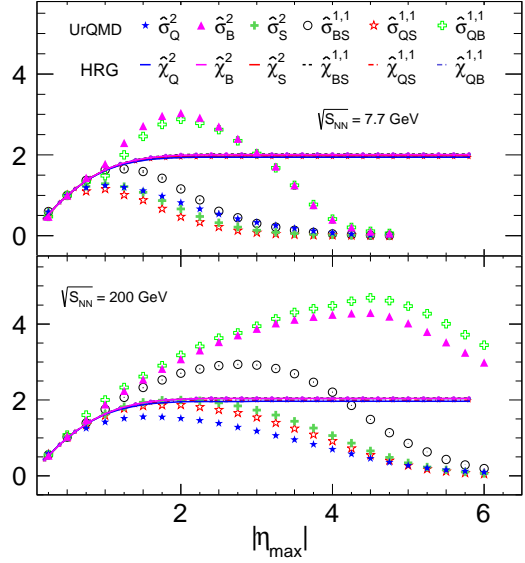}
\caption{(Color online) The ${\eta}_{\text{max}}$ dependence of all second order 
susceptibilities normalised by their values for ${p_T}_{\text{max}}=2$ GeV and 
$\eta_{\text{max}}=0.5$ in the HRG and UrQMD models.}
\label{fig.eta_all_onecanvas} 
\end{figure}

We have plotted the $\eta_{\text{max}}$ dependence of all the normalised susceptibilities with 
${p_T}_{\text{max}}=2$ GeV in Fig.~\ref{fig.eta_all_onecanvas}. They are all normalised 
by their values for $\eta_{\text{max}}=0.5$. Unlike the ${p_T}_{\text{max}}$ dependence, 
in this case all the plots collapse on each other and we don't see any mass ordering in 
the HRG results. The UrQMD results show a similar collapse for $\eta_{\text{max}}\leq1$ 
beyond which total charge conservation effect kicks in and the plots deviate from each 
other. While the mesonic moments peak around $\eta_{\text{max}}\sim1$ beyond which they 
start going to zero, the baryonic moments continue with their rise and peak at higher 
$\eta_{\text{max}}$. The location of the peaks of the baryonic moments are clearly 
dictated by the baryon stopping effect and closely follow the peak in
the $\eta$ distribution of baryons. The mesonic moment peaks are much less sensitive to \sNN.

\section{Summary}

The moments of conserved charges are important observables to probe the 
QCD phase diagram in heavy ion collisions. These provide glimpses of the thermal 
conditions prevalent in the fireball. They quantities are also expected to carry the signatures 
of non-thermal behaviour, like those close to the QCD critical point. Although the 
study of these observables is very well motivated, there are several experimental 
issues that need to be understood in order to interpret the data and draw physics 
conclusions. Using UrQMD and HRG models, we have studied some of the issues with respect to particle
identification as well as detector acceptances in the experiments
with regard to all the second order susceptibilities.
The main results of our study are as follows:
\begin{enumerate}
\item $\chi^2_Q$, $\chi^{11}_{BQ}$ and $\chi^{11}_{QS}$ can be measured accurately 
with the event-by-event measurements of the limited particle set: $\pi^\pm$, $K^\pm$, 
$p$ and $\bar{p}$, while the measured $\chi^2_B$ from net proton roughly scale as half 
of that expected from the complete particle set due to the missing neutrons,
\item We should have $|\eta_{\text{max}}|\leq0.5$ acceptance in order to have the ratio of conserved charge in bath to the total conserved charge from bath and system much smaller than half as it should be for grand canonical fluctuations for all \sNN$>10$ GeV,
\item Suitably normalised susceptibilities show a conserved charge ordering in the $p_T$ 
acceptance in HRG as well as in UrQMD. The net charge susceptibilities saturate to 
their maximum value at smaller ${p_T}_{\text{max}}$ value followed by net strangeness 
and net baryon. For a thermal medium, the ${p_T}_{max}$ dependence arises from the different 
masses of the relevant degrees of freedom that contribute to these conserved charge 
fluctuations. For a hadronic medium, this implies a clear ordering in the different 
conserved charges. By virtue of being normalised, they are independent of the fireball 
volume and thus can be reliably compared between experiments and theory. An experimental 
observation of such ordering will confirm the presence of the hadronic medium at the time 
of freeze-out of the susceptibilities. On the other hand, a negative result will hopefully 
lead to more interesting physics. It will be interesting in this context to study the 
influence of critical fluctuations on such ordering.

\end{enumerate}

\section{Acknowledgement}

We acknowledge many helpful discussions on susceptibilities with
Sourendu Gupta and Prithwish Tribedy. 
SC acknowledges 
XII$^{th}$ plan project no. 12-R$\&$D-NIS-5.11-0300 and CNT project PIC
XII-R$\&$D-VECC-5.02.0500 for support. NRS  is supported by the US
Department of Energy under Grant No. DE-FG02-07ER41485. This research
used resources of the LHC grid computing center at the Variable Energy
Cyclotron Center, Kolkata, India.

\section{Appendix: Statistical error estimation}

Let us consider the observable, which is the ratio of off-diagonal ($c_{1,1}$) to
diagonal ($c_{0,2} $) cumulants of conserved charged distributions,
\bea
C_{XY} = \alpha\frac{c_{1,1}}{c_{0,2}},
\eea
where $\alpha$ is a constant, and 
$X$ and $Y$ are the net charge ($Q$), net baryon ($B$) or
net strangeness ($S$). $C_{X,Y}$ can be expressed as
\bea
C_{X,Y} = \phi(c_{1,1},c_{0,2}).
\eea

Now using the error propagation formula, one can find the variance of
$\phi (X_{i},X_{j})$ as,
\begin{eqnarray}
V(\phi) &=& \sum_{i=1,j=1}^{n}{\frac{\partial\phi}{\partial X_{i}}\frac{\partial\phi}
{\partial X_{j}}}Cov(X_{i},X{j})  \nonumber  \\ 
&=& \sum_{i=1}^{n}{(\frac{\partial\phi}{\partial X_{i}})^{2}}V(X_{i})   \nonumber  \\ 
&&+\frac{1}{N}\sum_{i,j=1,i\neq j}^{n}{\frac{\partial\phi}  
{\partial X_{i}}\frac{\partial\phi}{\partial X_{j}}}Cov(X_{i},X{j}) 
\end{eqnarray}

Using $X_{1} = c_{1,1}$ and 
$X_{2} = c_{0,2}$, the variance of $\phi (c_{1,1},c_{0,2})$ can be
expressed in terms of the following:
\begin{eqnarray}
\label{var}
 V(\phi) &=& \ {(\frac{\partial\phi}{\partial c_{1,1}})^{2}}V(c_{1,1}) \ +  
\ {(\frac{\partial\phi}{\partial c_{0,2}})^{2}}V(c_{0,2})  \nonumber\\ 
&&+ \ 2{\frac{\partial\phi}{\partial c_{1,1}}\frac{\partial\phi}
{\partial c_{0,2}}}Cov(c_{1,1},c_{0,2})\ .
\end{eqnarray}
The general expression for covariance between moment ($c_{m,n}$ 
and $c_{k,l}$) ~\cite{Kendall} is
\bea
\label{cov}
Cov(c_{m,n},c_{k,l}) &=& \frac{1}{N} (c_{m+k,n+l} - c_{m,n}c_{k,l}).
\eea
Where $N$ is the total number of events in an ensemble. Using above two equations one obtains,
\bea
V(C_{X,Y}) &=& \frac{\alpha^{2}}{N}\ [ {(\frac{1}{c_{0,2}^{2}})}(c_{2,2} - 
c_{1,1}^{2}) \ +  \ {(\frac{c_{1,1}^{2}}{c_{0,2}^{4}})}(c_{0,4} - c_{0,2}^{2}) \ \nonumber\\
&&+ \ 2{(\frac{-c_{1,1}}{c_{0,2}^{3}})}(c_{1,3} - c_{1,1}c_{0,2})].
\eea
The error in $C_{X,Y}$ is finally written as
\bea
\text{error} = \sqrt{V(C_{XY})}.
\eea

Similarly, one can derive error of any observable (the higher order
off-diagonal and diagonal cumulants of different multiplicity distributions).

\end{document}